\documentclass[12pt]{article}

\usepackage[]{}
\usepackage{fontenc} 
\usepackage[english]{babel} 

\usepackage[hmarginratio=1:1,top=32mm, left=1.5cm]{geometry} 
\usepackage[hang, small,labelfont=bf,up,textfont=it,up]{caption} 
\usepackage{booktabs} 

\usepackage{lettrine} 
\usepackage{cite}
\usepackage{enumitem} 
\setlist[itemize]{noitemsep} 

\usepackage{abstract} 

\usepackage{titlesec} 
\renewcommand\thesection{\Roman{section}} 
\renewcommand\thesubsection{\Alph{subsection}} 
\titleformat{\section}[block]{\large\scshape\centering}{\thesection.}{1em}{} 
\titleformat{\subsection}[block]{\large}{\thesubsection.}{1em}{} 

\usepackage{titling} 
\usepackage{amsmath}
\usepackage{hyperref} 
\usepackage{cite}

\pretitle{\begin{center}\Huge\bfseries} 
\posttitle{\end{center}} 
\title{Novel way to the metric of higher dimensional
	rotating black holes } 
\author{
Amin Aghababaie Dastgerdi,$^{1}$\thanks{\href{mailto:aminaghababaie@ph.iut.ac.ir}{aminaghababaie@ph.iut.ac.ir}} \ \ Behrouz Mirza,$^{1}$\thanks{ \href{mailto:b.mirza@iut.ac.ir}{b.mirza@iut.ac.ir}} \ \ Naresh Dadhich$^{2}$\thanks{\href{mailto:nkd@iucaa.in}{nkd@iucaa.in}} \\
\normalsize \itshape{$^{1}$Department of Physics, Isfahan University of Technology, Isfahan 84156-83111, Iran} \\ 
\normalsize \itshape{$^{2}$ Inter University Centre for Astronomy \& Astrophysics, Post Bag 4, Pune 411007, India} \\[1ex] 
}
\date{} 
\renewcommand{\maketitlehookd}{%
\begin{abstract}
\normalfont {We wish to carry forward to higher dimensions the insightful and novel method of obtaining the Kerr
	metric proposed by one of us [Gen. Relativ. Gravit. 45, 2383 (2013)] for deriving the Myers-Perry rotating
	black hole metric. We begin with a flat spacetime metric written in oblate spheroidal coordinates
	(ellipsoidal geometry) appropriate for the inclusion of rotation, and then introduce arbitrary functions to
	introduce a gravitational potential due to mass, which are then determined by requiring that a massless
	particle experiences no acceleration, while a massive particle feels Newtonian acceleration at large $r$.
	We further generalize the method to include the cosmological constant $\Lambda$ to obtain the Myers–Perry–
	de Sitter/anti–de Sitter black hole metric.}
\end{abstract}
}

\begin{document}

\maketitle


\section{{\bfseries Introduction}}
Einstein's gravitational equations are highly nonlinear differential equations and hence are difficult to solve. However, the first exact solution describing the field of a static mass point was obtained by  Schwarzschild \cite{schw} immediately after the theory was discovered. It then took 48 years for the discovery of a rotating black hole solution by  Kerr \cite{Kerr}. The charged versions of the Schwarzschild and Kerr solutions followed  soon afterwards (Refs. \cite{Reiss,Nord} and \cite{kerrnewman1,kerrnewman2} respectively). \\

However, it was easy to find higher-dimensional versions of the static solution \cite{tangherlini}. Myers and Perry obtained the higher-dimensional generalization of the rotating Kerr metric \cite{MP}. However, there exists no higher-dimensional exact solution of the Einstein-Maxwell equations describing a charged rotating black hole, i.e, no higher-dimensional analog of a Kerr-Newman black hole has yet been found. Because of the nonlinearity and complexity of the equations, there are only a handful of physically meaningful exact solutions. \\

With this background, it is important to devise ingenious techniques and methods where one does not have to solve the Einstein equations to obtain the required metric. One such technique is the Newman-Janis (NJ) algorithm \cite{NJ} that converts static a Schwarzschild solution into a rotating Kerr solution by applying a simple complex transformation. As a matter of fact, the Kerr-Newman metric of a charged rotating black hole \cite{kerrnewman1,kerrnewman2} was also obtained by employing this technique. The NJ algorithm has been widely employed in various cases, in particular for obtaining a rotating NUT black hole in \cite{Erbin}, a rotating black hole with a scalar field \cite{NJ1} and in modified gravity \cite{NJ2}. A generalized formulation of the Newman-Janis algorithm for five-dimensional black holes with arbitrary angular momenta and in seven dimensions  with equal angular momenta was recently proposed in Refs.
\cite{Mirza5D,MirzaOdd}. It is a prescription that works like magic but one knows very little about why and how it works. This question is addressed to some extent in Ref. \cite{NJ3} where the uniqueness of the NJ algorithm was established. At any rate, the physical reason why it works is not so obvious and remains unclear. \\

Recently, Dadhich proposed an insightful method
of obtaining black hole metrics, first for the static
Schwarzschild \cite{DadhichSchw} and then for the rotating Kerr black
hole \cite{Dadhich}. The most attractive feature of the method is that it is driven by simple physical and geometrical considerations, and hence one clearly can understand and see why and how it works. First, a spatial geometry appropriate for the gravitating object is chosen in flat spacetime: spherical for the static nonrotating case and ellipsoidal for the rotating case. Then, two arbitrary functions are introduced to generate gravitational potential due to mass. Finally, these functions are determined by appealing to two physical guiding principles: massless particles experience no acceleration while massive particles experience Newtonian acceleration in the first approximation. These conditions are based on the principle that in vacuum a photon should experience no acceleration to maintain the constant velocity of light, while Newton's law should always be included in general relativity \cite{DadhichSchw} in the first approximation. The physical motivation of this method is therefore clear and transparent. \\

In this paper, we wish to take Dadhich's prescription \cite{Dadhich} to higher dimensions to obtain the Myers-Perry metric of a rotating black hole. The paper is organized as follows. In the next section we apply the method to obtain the Myers-Perry metric in both odd ($d=2n+1$) and even ($d=2n+2$) dimensions, where $n=[(d-1)/2]$  is the maximum number of rotation parameters a black hole can have in dimension $d$. In Sec. III, we include  the cosmological constant $\Lambda$ and obtain  the Myers-Perry-anti de Sitter/de Sitter (AdS/dS) metric in four and five dimensions. This is followed by a consideration of a general Myers-Perry-AdS metric in arbitrary dimensions. We end with a discussion.


\section{{\bfseries Higher-dimensional Myers-Perry rotating black hole metrics}}\label{sec:one}
We shall first obtain the Myers-Perry metric for a five-dimensional rotating black hole to illustrate the method, and then apply it to arbitrary higher-dimensional black holes with $[(d-1)/2]$ rotation parameters. We begin by writing the flat Minkowski metric in the ellipsoidal form, which is appropriate for the  inclusion of rotation. To introduce a gravitational potential, two arbitrary functions of $r$ are added to the metric without disturbing axial symmetry. These functions would be determined by requiring that axially falling massive particles experience Newtonian acceleration for large $r$ and massless ones experience no acceleration. This will finally lead to the required Myers-Perry metric.
\subsection{\bfseries Five-dimensional Myers-Perry black hole }
Let us begin with the Minkowski metric in Cartesian coordinates in five dimensions,
\begin{equation}
	ds^2=dt^2-\sum_{i=1}^{2}{\left({{dx}_i}^2+{{dy}_i}^2\right)}. \label{eq:ds2m}
\end{equation}
We transform it to the oblate spheroidal coordinates by the following transformations:
\begin{equation}
\begin{aligned}
&x_1=\sqrt{r^2+a^2} \ \sin{\vartheta}\cos{\varphi},\\
&y_1=\sqrt{r^2+a^2} \ \sin{\vartheta}\sin{\varphi},\\
&x_2=\sqrt{r^2+b^2} \ \sin{\vartheta}\cos{\psi},\\
&y_2=\sqrt{r^2+b^2} \ \sin{\vartheta}\sin{\psi}.
	\end{aligned}
\end{equation}
The above $d=5$ Minkowski metric in Eq. \eqref{eq:ds2m} takes the form
\begin{equation}
	{ds}^2={dt}^2-\frac{r^2\rho^2}{\left(r^2+a^2\right) \left(r^2+b^2\right)}\ dr^2-\rho^2{d\vartheta}^2-\left(r^2+a^2\right)\ {\sin}^2\vartheta\ d\varphi^2  -\left(r^2+b^2\right)\ {\cos}^2\vartheta d\psi^2,\label{eq:dsos}
\end{equation}
where $\rho^2=r^2+a^2 \cos^2{\theta}+b^2\sin^2{\theta} $, and $a$ and $b$ are arbitrary constants.  It could be further  transformed into the Boyer-Lindquist form \cite{Boyer}, 
\begin{equation}
	\begin{aligned}
&ds^2=\frac{\Delta_0}{\rho^2}{d\tau}^2-\frac{\rho^2}{\Delta_0}dr^2-\rho^2{d\vartheta^2}-\frac{\sin^2{\theta}}{\rho^2}\left[ ( r^2+a^2) d\varphi-a \ dt\right] ^2\\
&-\frac{cos^2\vartheta}{\rho^2}\ [\left(r^2+b^2\right)d\psi-b\ dt]^2\\
&-\frac{1}{r^2\rho^2}\ [ab\ dt-b\ sin^2\vartheta\left(r^2+a^2\right)d\varphi-a\ cos^2\vartheta\left(r^2+b^2\right)d\psi]^2,
\end{aligned}\label{eq:dsdelta}
\end{equation}
where $d\tau=dt-asin^2\vartheta\ d\varphi-bcos^2\vartheta\ d\psi\,
\Delta_0=\frac{1}{r^2}\left(r^2+a^2\right)\ \left(r^2+b^2\right) \, .
$ 
Clearly, the gravitational potential which would be a function of $r$ only would appear in $\Delta_0$. The rest of the metric remains unaltered as above adhering to axial symmetry, and so we write
\begin{equation}
	\begin{aligned}
		&ds^2=\frac{f(r)}{\rho^2}{d\tau}^2-\frac{\rho^2}{g(r)}dr^2-\rho^2{d\vartheta^2}-\frac{\sin^2{\theta}}{\rho^2}\left[ ( r^2+a^2) d\varphi-a \ dt\right] ^2\\
	&-\frac{cos^2\vartheta}{\rho^2}\ [\left(r^2+b^2\right)d\psi-b\ dt]^2\\
	&-\frac{1}{r^2\rho^2}\ [ab\ dt-b\ sin^2\vartheta\left(r^2+a^2\right)d\varphi-a\ cos^2\vartheta\left(r^2+b^2\right)d\psi]^2,
	\end{aligned}	\label{eq:dsfg}
	\end{equation}
where $f(r)$ and $g(r)$ are arbitrary functions to be determined. Asymptotically, the metric should become the Minkowski metric, and hence as $r\to\infty$, $f\left(r\right)=g\left(r\right)= \Delta_0 = \frac{1}{r^2}\left(r^2+a^2\right)\left(r^2+b^2\right)$. There are two options for investigating radial motion: the subspace $\vartheta=0$ with $\psi=const$ and $b=0$, and the subspace $\vartheta=\frac{\pi}{2}$ with $\varphi=const$ and $a=0$. We will opt for the former, i.e., axially falling particles in subspace $\vartheta=0$ and $\psi=const$  with only one rotating parameter $\left(b=0\right)$. There is no loss of generality because $f(r)$ and $g(r)$ would involve a gravitational potential, which is not going to depend on whether the black hole has one or two rotations. The Lagrangian for particle motion in this subspace is given by 
\begin{equation}
\mathcal{L}=g_{\mu\nu}{\dot{x}}^\mu{\dot{x}}^\nu=\frac{f}{\left(r^2+a^2\right)}{\dot{t}}^2-\frac{\ \left(r^2+a^2\right)}{g}{\dot{r}}^2=\mu^2,\label{eq:energy5d}
\end{equation}
where a dot denotes a partial derivative with respect to an affine parameter $\lambda$, and $\mu$ is the mass of the particle. Since the Lagrangian is free of $t$, the corresponding canonical momentum is conserved to give
\begin{equation}
	p_t=\frac{f}{\left(r^2+a^2\right)}\dot{t}=E.
\end{equation}
Putting this back into the Eq. (\ref{eq:energy5d}), we get
\begin{equation}	{\dot{r}}^2=\left(\frac{\left(r^2+a^2\right)}{f}E^2-\mu^2\right)\frac{g}{\left(r^2+a^2\right)}\ .\label{eq:Motionfg}
\end{equation}
Since massless particles $\left(\mu=0\right)$ experience no acceleration, $(\ddot{r}=0)$,
\begin{equation}
\ddot{r}=\frac{E^2}{2}\ \left(\frac{g}{f}\right)^\prime=0,
\end{equation}
where a prime denotes a partial derivative with respect to $r$ $(\frac{d}{d\lambda}=\frac{dr}{d\lambda}\frac{d}{dr}=\dot{r}\frac{d}{dr})$. Thus, we have
\begin{equation}
	\begin{aligned}
	\frac{g}{f}=Const. = 1.
	\end{aligned}
\end{equation}
The above constant should be set equal to $1$ as this is the only case where the metric (\ref{eq:dsfg}) tends to the flat Minkowski metric as $r\rightarrow\infty$, i.e, it is asymptotically flat. We can now rewrite Eq. (\ref{eq:Motionfg}) for massive particle with $\mu=1$ as
\begin{equation}
	{\dot{r}}^2=E^2-\frac{f}{r^2+a^2} \ .\ \ \ (\ \mu=1)\label{eq:Motionff}
\end{equation}
Upon differentiating we get
\begin{equation}
\ddot{r}=-\frac{1}{2}\left[\frac{f^\prime}{r^2+a^2}-\frac{2rf}{(r^2+a^2)^2}\right]\ .\label{eq:accmass}
\end{equation}
Now, taking the asymptotic flatness condition into account, we write $f\left(r\right)=g\left(r\right)=\frac{1}{r^2}\left(r^2+a^2\right)\left(r^2+b^2\right)+\psi(r)$ where the function $\psi\left(r\right) \to 0$ as $r\to\infty$. Replacing $f\left(r\right)$ in Eq. (\ref{eq:accmass}) yields
\begin{equation}
\ddot{r}=-\frac{1}{2}\left[\frac{2r+\psi^\prime}{r^2+a^2}-\frac{2r(r^2+a^2+\psi)}{(r^2+a^2)^2}\right].
\end{equation}
which for large $r$ reduces to
\begin{equation}
\ddot{r}=\left[\frac{\psi}{r^3}-\frac{\psi^\prime}{{2r}^2}\right].\label{eq:accpsi}
\end{equation}
This should now agree with the Newtonian acceleration in $d=5$ (note that the potential in dimension $d$ goes as $M/r^{d-3}$)\cite{NewtPot}, and thus $\ddot{r}=-\frac{2M}{r^3}$, giving the equation
\begin{equation}
r\psi' - 2\psi - 4M=0\, .
\end{equation}
This is readily integrated to give
\begin{equation}
\psi\left(r\right)\ = -2M\ ,
\end{equation}
where the integration constant is set to zero for asymptotic flatness. We would have arrived at the same result had we followed the other alternative, i.e., the subspace $\vartheta=\frac{\pi}{2}$ with $\varphi=const$ and $a=0$.

Plugging in the other rotation parameter and the functions $f\left(r\right)$ and $g\left(r\right)$, we write the general form of the five-dimensional Myers-Perry metric for a rotating black hole of mass $M$ and rotation parameters $a$ and $b$ as follows:
\begin{equation}
	\begin{aligned}
	&{ds}^2=\frac{\Delta}{\rho^2}d\tau^2-\frac{\rho^2}{\Delta}\ {dr}^2-\rho^2d\vartheta^2-\frac{sin^2\vartheta}{\rho^2}\ [\left(r^2+a^2\right)d\varphi-a\ dt]^2\\
	&-\frac{cos^2\vartheta}{\rho^2}\ [\left(r^2+b^2\right)d\psi-b\ dt]^2\\
	&-\frac{1}{r^2\rho^2}\ [ab\ dt-b\ sin^2\vartheta\left(r^2+a^2\right)d\varphi-a\ cos^2\vartheta\left(r^2+b^2\right)d\psi]^2,
	\end{aligned} \label{eq:MP5D1}
\end{equation}
where  $\Delta=f\left(r\right)=g\left(r\right)=\frac{1}{r^2}\left(r^2+a^2\right)\left(r^2+b^2\right)-2M$. The metric (\ref{eq:MP5D1}) can also be written in the following Myers-Perry form:
\begin{equation}
	\begin{aligned}
&{ds}^2={dt}^2-\frac{2M}{\rho^2}(dt+a\ {sin}^2\vartheta\ d\varphi+b\ {cos}^2\vartheta\ d\psi)^2-\rho^2{d\vartheta}^2 -\frac{r^2\rho^2}{\left(r^2+a^2\right)\ \left(r^2+b^2\right)-2Mr^2}{\ dr}^2\\
&-\left(r^2+a^2\right){\ sin}^2\vartheta\ d\varphi^2-\left(r^2+b^2\right)\ {cos}^2\vartheta\ d\psi^2.
	\end{aligned}
\end{equation}\\

We would like to emphasize that in the Minkowski metric in ellipsoidal form, we have only replaced $\Delta_0$ by the free functions $f(r)$ and $g(r)$ which were determined by appealing to physically motivated conditions, i.e., massive particles experience Newtonian acceleration and no acceleration for massless ones. Nothing else was added from outside by hand. In principle it would be possible to carry on with the same procedure for $d>5$; however, it would be too involved and cumbersome to implement. On the other hand, the above Myers-Perry form is ideally suited for this purpose.

\subsection{\bfseries Odd dimensions $(d=2n+1)$ }
The metric of an odd $(2n+1)-$dimensional flat spacetime can be written as follows:
\begin{equation}
	{ds}^2={dt}^2-\sum_{i=1}^{n}{\ \left({{dx}_i}^2+{{dy}_i}^2\right)}.\label{eq:MInodd}
\end{equation}
Now we transform the metric (\ref{eq:MInodd}) to the required form by employing the following transformations:
\begin{equation}
	\begin{aligned}
&x_i=\sqrt{r^2+{a_i}^2}\ \ \mu_i\ \cos{\varphi_i},\\
&y_i=\sqrt{r^2+{a_i}^2}\ \ \mu_i \ \sin{\varphi_i},
	\end{aligned}
\end{equation}
with
\begin{equation}
\sum_{i=1}^{n}{\mu_i}^2=1,
\end{equation}
where $\mu_i$ are direction cosines. Now the metric takes the form 
\begin{equation}
ds^2=dt^2-\frac{\Pi F}{\Pi}{dr}^2-\sum_{i=1}^{n}\left(\ r^2+a_i^2\right)\left(d\mu_i^2+\mu_i^2{d\varphi}_i^2\right),\label{eq:oddbimass}
\end{equation}
where
\begin{equation}
	\begin{aligned}
&F=1-\sum_{i=1}^{n}\frac{a_i^2\mu_i^2}{r^2+a_i^2}=r^2\sum_{i=1}^{n}\frac{\mu_i^2}{r^2+a_i^2},\\
&\Pi=\prod_{i=1}^{n}{(r^2}+a_i^2).
	\end{aligned}
\end{equation}
This is a flat metric in the Myers-Perry form; how do we inject a gravitational potential due to a point mass into it? In view of the Myers-Perry metric (\ref{eq:MP5D1}), we have to add a term $d\tau^2$ with an appropriate coefficient to introduce rotation. Thus, we write by inserting as before the functions $f(r)$ and $g(r)$,
\begin{equation}
ds^2=dt^2-\frac{f\left(r\right)}{\Pi F}(dt+\sum_{i=1}^{n}a_i\mu_i^2{d\varphi}_i)^2-\frac{\Pi F}{g\left(r\right)}{dr}^2-\sum_{i=1}^{n}(r^2+a_i^2)\left(d\mu_i^2+\mu_i^2{d\varphi}_i^2\right).\label{eq:oddfg}
\end{equation}
Here we should note that  $\frac{f\left(r\right)}{\Pi F}$  must tend to zero asymptotically as $r\rightarrow\infty$. As before we would like to investigate the radial motion and so we set $\mu_1=\mu_2=\ldots=\mu_{n-1}=0$, $a_n=0$ and $\varphi_n=Const$. Therefore, we have $\mu_n=1$ and $F=1$, and the metric (\ref{eq:oddfg}) takes the following form:
\begin{equation}
ds^2=dt^2-\frac{f\left(r\right)}{\Pi}dt^2-\frac{\Pi}{g\left(r\right)}{dr}^2.
\end{equation}
Retracing the same steps as before we readily arrive at
\begin{equation}
{\dot{r}}^2=\left[E^2\left(\frac{\Pi}{\Pi-f}\right)-\mu^2\right]\ \frac{g}{\Pi}.\label{eq:motionoddfg}
\end{equation}
Now for $\ddot{r}=0$ for massless particles $(\mu=0)$, we have $(\frac{d}{d\lambda}=\frac{dr}{d\lambda}\frac{d}{dr}=\dot{r}\frac{d}{dr})$
\begin{equation}
\ddot{r}=\frac{E^2}{2}(\frac{g}{\Pi-f})^\prime=0\Longrightarrow\frac{g}{\Pi-f}=1,\label{eq:photonodd}
\end{equation}
with the constant being set to unity for asymptotic flatness. By replacing $g\left(r\right)=\Pi-f(r)$ in Eq. (\ref{eq:motionoddfg}) and assuming $\mu=1$, we write
\begin{equation}
{\dot{r}}^2=E^2-1+\frac{f}{\Pi}\ \Longrightarrow\ddot{r}=\frac{1}{2}\left(\frac{f^\prime}{\Pi}-\frac{f\ \Pi^\prime}{\Pi^2}\right).
\end{equation}
We know that asymptotically $\Pi \sim r^{2n}$, so at very large $r$, $\ddot{r}$ approximates to
\begin{equation}
	\ddot{r}=\frac{f^\prime}{{2r}^{2n}}-\frac{nf}{r^{2n+1}}\ , \label{eq:accodd}
\end{equation}
which should agree with the Newtonian acceleration, $(3-d)M/r^{d-2}$. This leads to the equation
\begin{equation}
	rf' -2nf + 2(n-1)Mr^2 =0,\label{eq:newtond}
\end{equation}
for $d=2n+1$. Respecting asymptotic flatness, it can be integrated to give
\begin{equation}
f\left(r\right)=2Mr^2.,
\end{equation}
and from Eq. (\ref{eq:photonodd}) we write
\begin{equation}
g\left(r\right)=\Pi-2Mr^2.
\end{equation}
Upon replacing $f(r)$ and $g(r)$ in Eq. (\ref{eq:oddfg}), we arrive at the Myers-Perry metric in odd $d=2n+1$ dimensions as follows
\begin{equation}
ds^2=dt^2-\frac{2Mr^2}{\Pi\ F}(dt+\sum_{i=1}^{n}{\ a_i}{\ \mu}_i^2{\ d\varphi}_i)^2-\frac{\Pi\ F}{\Pi-2Mr^2}{dr}^2-\sum_{i=1}^{n}{\ (}r^2+a_i^2)\left(d\mu_i^2+\mu_i^2{\ d\varphi}_i^2\right).
\end{equation}
\subsection{\bfseries Even dimensions $(d=2n+2)$}
In $d=2n+2$, we begin with
\begin{equation}
{ds}^2={dt}^2-\sum_{i=1}^{n}{\ \left({{dx}_i}^2+{{dy}_i}^2\right)}-{dz}^2,
\end{equation}
with the transformations
\begin{equation}
	\begin{aligned}
		&x_i=\sqrt{r^2+{a_i}^2}\ \ \mu_i\ \cos{\varphi_i},\\
		&y_i=\sqrt{r^2+{a_i}^2}\ \ \mu_i \ \sin{\varphi_i},\\
		&z=r\ \alpha,
	\end{aligned}
\end{equation}
if we write
\begin{equation}
\sum_{i=1}^{n}{\mu_i}^2+\alpha^2=1,
\end{equation}
the metric takes the form
\begin{equation}
ds^2=dt^2-\frac{\Pi\ F}{\Pi}{dr}^2-\sum_{i=1}^{n}{\ \left(\ r^2+a_i^2\right)\left(d\mu_i^2+\mu_i^2\ {d\varphi}_i^2\right)}-r^2{d\alpha}^2,
\end{equation}
where $F$ and $\Pi$ are as given above in Eq. (23).

Retracing all of the steps as in the odd $(d=2n+1)$ case above, we obtain
\begin{equation}
	f\left(r\right)=2Mr \, , \ g\left(r\right)=\Pi-2Mr.
\end{equation}
Now the metric for an even $d=2n+2$-dimensional rotating black hole is writas follows:
\begin{equation}
	ds^2=dt^2-\frac{2Mr}{\Pi F}\left(dt+\sum_{i=1}^{n}a_i\mu_i^2{d\varphi}_i)^2-\frac{\Pi F}{\Pi-2Mr}{dr}^2-\sum_{i=1}^{n}(r^2+a_i^2\right)\left(d\mu_i^2+\mu_i^2{d\varphi}_i^2\right)-r^2{d\alpha}^2.ten 
\end{equation}
\section{\bfseries Myers-Perry-AdS black hole}
In this section, we apply the method to include $\Lambda$ and obtain rotating black holes in AdS spacetime. We first obtain the Kerr-AdS metric in four dimensions, and then obtain the five-dimensional Myers-Perry-AdS black hole metric.\\

Let us begin with the AdS metric in the usual spherical coordinates,
\begin{equation}
	ds^2=\left(1+\frac{r^2}{l^2}\right)dt^2-\frac{{dr}^2}{1+\frac{r^2}{l^2}}-r^2\ {d\vartheta}^2-r^2\ {sin}^2\vartheta\ d\varphi^2.\label{eq:AdsSPhe}
\end{equation}
if we write $r=\sqrt{{x_1}^2+{x_2}^2+{x_3}^2}$, the metric takes the form 
\begin{equation}
	ds^2=\left(1+\frac{{x_1}^2+{x_2}^2+{x_3}^2}{l^2}\right)dt^2-{{dx}_1}^2-{{dx}_2}^2-d{x_3}^2 +\frac{\frac{(x_1\ dx_1+x_2\ dx_2+{x\ }_3dx_3)^2}{l^2}}{1+\frac{{x_1}^2+{x_2}^2+{x_3}^2}{l^2}}.\label{eq:ADSDEK}
\end{equation}
Now we go to oblate spheroidal coordinates by the following  coordinate transformations,
\begin{equation}
	\begin{aligned}
		&x_1=\sqrt{\frac{r^2+a^2}{1-\frac{a^2}{l^2}}}\ \sin{\vartheta} \ \cos{\left[\varphi+\frac{a}{l^2}\ t\right]},\\
		&x_2=\sqrt{\frac{r^2+a^2}{1-\frac{a^2}{l^2}}}\ \sin{\vartheta} \ \sin{\left[\varphi+\frac{a}{l^2}\ t\right]},\\
		&x_3=r\ \cos{\vartheta}.
	\end{aligned}\label{ADSTRANS}
\end{equation}
Then, the metric takes the following oblate spheroidal form,
\begin{equation}
	{ds}^2=\frac{\Delta}{\rho^2}\left[dt-\frac{a \sin^2 \vartheta}{\Xi}d\varphi\right] ^2-\frac{\rho^2}{\Delta} {dr}^2-\frac{\rho^2}{\Delta_\vartheta}{d\vartheta}^2-\frac{\Delta_\vartheta \sin^2 \vartheta }{\rho^2}\left[a dt-\frac{r^2+a^2}{\Xi} d\varphi\right] ^2,\label{eq:ADSOSBIMASS}
\end{equation}
where
\begin{equation}
	\begin{aligned}
		&\rho^2=r^2+a^2cos^2\vartheta,\\
		&\Xi=1-\frac{a^2}{l^2},\\
		&\Delta=(r^2+a^2)(1+\frac{r^2}{l^2}),\\
		&\Delta_\vartheta=1-\frac{a^2}{l^2} \cos^2\vartheta.
	\end{aligned}
\end{equation}
To introduce a gravitational potential due to mass, as before we introduce the functions $f(r)$ and $g(r)$ into Eq. (\ref{eq:ADSOSBIMASS})
\begin{equation}
	{ds}^2=\frac{f(r)}{\rho^2}\left[dt-\frac{a \sin^2 \vartheta}{\Xi}d\varphi\right] ^2-\frac{\rho^2}{g(r)} {dr}^2-\frac{\rho^2}{\Delta_\vartheta}{d\vartheta}^2-\frac{\Delta_\vartheta \sin^2 \vartheta }{\rho^2}\left[a dt-\frac{r^2+a^2}{\Xi} d\varphi\right] ^2\ .\label{eq:ADSOSfg}
\end{equation}
Now we consider motion along the axis of rotation $(\vartheta=0)$ of massive and massless particles for which the Lagrangian takes the form
\begin{equation}
	\mathcal{L}=\frac{f\left(r\right)}{\rho^2}\ {\dot{t}}^2-\frac{\rho^2}{g\left(r\right)}\ {\dot{r}}^2=\mu^2.
\end{equation}
As before, ${E=p}_t=\frac{f\left(r\right)}{\rho^2}\dot{t}$ and we write
\begin{equation}
	{\dot{r}}^2=\left(\frac{\rho^2}{f\left(r\right)}\ E^2-\mu^2\right)\frac{g\left(r\right)}{\rho^2}\ ,
\end{equation}
Now we implement the two guiding conditions. For massless $(\mu=0)$ particles, $\ddot{r}=0$, which readily leads to $f\left(r\right)=g\left(r\right)$. Then, for massive particles with $\mu=1$, we have
\begin{equation}
	\ddot{r}=-\frac{1}{2}\left[\frac{f^\prime}{r^2+a^2}-\frac{2rf}{(r^2+a^2)^2}\right].
\end{equation}
Let us write $f\left(r\right)=g(r)=\left(r^2+a^2\right)\left(1+\frac{r^2}{l^2}\right)+k(r)$, where the function $k(r)$ goes to zero asymptotically as $r\rightarrow\infty$. We substitute this into the above  acceleration equation, which at large $r$ should be equal to that of the Schwarzschild-AdS \cite{NewtLamb}, and hence we write
\begin{equation}
	\ddot{r}=-\frac{1}{2}\left[\frac{2r}{l^2}+\frac{k^\prime}{r^2}-\frac{2k}{r^3}\right]=-\frac{M}{r^2}-\frac{r}{l^2}.
\end{equation}
This leads to the equation
\begin{equation}
	k' -2k/r - 2M=0
\end{equation}
which is readily solved to give
\begin{equation}
	k = -2Mr\ .
\end{equation}
The constant of integration has been set to zero for the same reason as before. Finally, we have $f\left(r\right)=g\left(r\right)=\left(r^2+a^2\right)\left(1+\frac{r^2}{l^2}\right)-2Mr$, and by replacing $f(r)$ and $g\left(r\right)$ in Eq. (\ref{eq:ADSOSfg}) we obtain the Kerr-AdS metric \cite{Ads1,Ads2,Ads3,Ads4} in Boyer-Lindquist coordinates:
\begin{equation}
	{ds}^2=\frac{\Delta_r}{\rho^2}\left[dt-\frac{a \sin^2 \vartheta}{\Xi}d\varphi\right] ^2-\frac{\rho^2}{\Delta_r} {dr}^2-\frac{\rho^2}{\Delta_\vartheta}{d\vartheta}^2-\frac{\Delta_\vartheta \sin^2 \vartheta }{\rho^2}\left[a dt-\frac{r^2+a^2}{\Xi} d\varphi\right] ^2,
\end{equation}
where
\begin{equation}
	\Delta_r=\left(r^2+a^2\right)\left(1+\frac{r^2}{l^2}\right)-2Mr.
\end{equation}\\

Now we obtain the five-dimensional rotating black hole metric in AdS spacetime. We begin similarly with the AdS metric in oblate spheroidal form and perform the transformations
\begin{equation}
	\begin{aligned}
		&x_1=\sqrt{\frac{r^2+a^2}{1-\frac{a^2}{l^2}}}\ \sin{\vartheta}\cos{\left[\varphi+\frac{a}{l^2}\ t\right]},\\
		&y_1=\sqrt{\frac{r^2+a^2}{1-\frac{a^2}{l^2}}}\ \sin{\vartheta}\sin{\left[\varphi+\frac{a}{l^2}\ t\right]},\\
		&x_2=\sqrt{\frac{r^2+b^2}{1-\frac{b^2}{l^2}}}\cos{\vartheta}\cos{\left[\psi+\frac{b}{l^2}t\right]},\\
		&y_2=\sqrt{\frac{r^2+b^2}{1-\frac{b^2}{l^2}}}\ \cos{\vartheta}\sin{\left[\psi+\frac{b}{l^2}\ t\right]
		},
	\end{aligned}
\end{equation}
to get to the required spheroidal form,
\begin{equation}
	\begin{aligned}
		&{ds}^2=\frac{\Delta}{\rho^2}\left[dt-\frac{a \sin^2 \vartheta}{\Xi_a}d\varphi-\frac{b \cos^2 \vartheta}{\Xi_b}d\psi\right] ^2-\frac{\rho^2}{\Delta} {dr}^2-\frac{\rho^2}{\Delta_\vartheta}{d\vartheta}^2\\
		&-\frac{\Delta_\vartheta \sin^2 \vartheta }{\rho^2}\left[a dt-\frac{r^2+a^2}{\Xi_a} d\varphi\right]^2-\frac{\Delta_\vartheta \cos^2 \vartheta }{\rho^2}\left[b dt-\frac{r^2+b^2}{\Xi_b} d\psi\right]^2\\
		&-\frac{\left(1+\frac{r^2}{l^2}\right)}{r^2\rho^2}(ab\ dt-\frac{b\left(r^2+a^2\right)sin^2\vartheta}{\Xi_a}d\varphi\ -\frac{a\left(r^2+b^2\right)cos^2\vartheta}{\Xi_b}\ d\psi)^2,
	\end{aligned}
\end{equation}
where
\begin{equation}
	\begin{aligned}
		&\rho^2=r^2+a^2cos^2\vartheta+b^2sin^2\vartheta,\\
		&\Xi_a=1-\frac{a^2}{l^2},\\
		&\Xi_b=1-\frac{b^2}{l^2},\\
		&\Delta=\frac{1}{r^2}(r^2+a^2)(r^2+b^2)(1+\frac{r^2}{l^2}),\\
		&\Delta_\vartheta=1-\frac{a^2}{l^2} \cos^2\vartheta-\frac{b^2}{l^2} \sin^2\vartheta.
	\end{aligned}
\end{equation}
Proceeding in the same way as before, we have $f(r)=g(r)$ and write
\begin{equation}
 f\left(r\right)=g(r)=\frac{1}{r^2}\left(r^2+a^2\right)\left(r^2+b^2\right)\left(1+\frac{r^2}{l^2}\right)+k(r).
\end{equation}
The corresponding acceleration equation in this case would be
\begin{equation}
	\ddot{r}=-\frac{1}{2}\left[\frac{2r}{l^2}+\frac{k^\prime}{r^2}-\frac{2k}{r^3}\right] =\ -\frac{2M}{r^3}-\frac{r}{l^2}.
\end{equation}
So the equation to be solved becomes
\begin{equation}
	k'-2k/r = 4M/r
\end{equation}
which can be solved to give $k=-2M$.\\

The required five-dimensional metric of a rotating AdS black hole is then given by \cite{Hawking}
\begin{equation}
	\begin{aligned}
		&{ds}^2=\frac{\Delta_r}{\rho^2}\left[dt-\frac{a \sin^2 \vartheta}{\Xi_a}d\varphi-\frac{b \cos^2 \vartheta}{\Xi_b}d\psi\right] ^2-\frac{\rho^2}{\Delta_r} {dr}^2-\frac{\rho^2}{\Delta_\vartheta}{d\vartheta}^2\\
		&-\frac{\Delta_\vartheta \sin^2 \vartheta }{\rho^2}\left[a dt-\frac{r^2+a^2}{\Xi_a} d\varphi\right]^2-\frac{\Delta_\vartheta \cos^2 \vartheta }{\rho^2}\left[b dt-\frac{r^2+b^2}{\Xi_b} d\psi\right]^2\\
		&-\frac{\left(1+\frac{r^2}{l^2}\right)}{r^2\rho^2}(ab\ dt-\frac{b\left(r^2+a^2\right)sin^2\vartheta}{\Xi_a}d\varphi\ -\frac{a\left(r^2+b^2\right)cos^2\vartheta}{\Xi_b}\ d\psi)^2,
	\end{aligned}
\end{equation}
where we have $\Delta_r=\frac{1}{r^2}\left(r^2+a^2\right)\left(r^2+b^2\right)\left(1+\frac{r^2}{l^2}\right)-2M$.

By writing $1/l^2 \rightarrow -1/l^2$ in the above we could similarly obtain the dS version of rotating black holes. \\

\section{\bfseries General Myers-Perry-AdS Metric in arbitrary dimension}

First, we consider the AdS metric in Cartesian coordinates for odd dimensions $(d=2n+1)$, 
\begin{equation}
	ds^2=-\left(1+\frac{\sum_{i=1}^{n}({x}_i^2+{y}_i^2)}{l^2}\right) dt^{\prime2}+\sum_{i=1}^{n} ({dx}_i^2+{dy}_i^2)-\frac{\left(\sum_{i=1}^{n} ({x}_i dx_i+{y}_i dy_i)\right)^2}{l^2+\sum_{i=1}^{n} ({x}_{i}^{2}+{y}_{i}^{2})}\ .
\end{equation}
We employ the new transformations
\begin{equation}
	\begin{aligned}
		&x_i=\sqrt{\frac{(r^2+a_i^2)}{1-\frac{a_i^2}{l^2}}} \ \mu_i \  cos[(1-\frac{a_i^2}{l^2})\varphi_i]\\
		&y_i=\sqrt{\frac{(r^2+a_i^2)}{1-\frac{a_i^2}{l^2}}} \ \mu_i \  sin[(1-\frac{a_i^2}{l^2})\varphi_i]\\
		&t^{\prime}=\prod_{i=1}^{n}(1-\frac{a_i^2}{l^2})t,
	\end{aligned} \label{eq:ADSTRANSF}
\end{equation}
where
\begin{equation}
	\sum_{i=1}^{n}{\mu_i}^2=1,
\end{equation}
and the parametrization for $\mu_i$ is \cite{Kerr-ADS}
\begin{equation} \mu_i^2=\frac{\prod_{\alpha=1}^{n-1}(a_i^2-y_\alpha^2)}{\prod_{k=1}^{\prime n}(a_i^2-a_k^2)}\, . \label{eq:ODDyalpha}
\end{equation}
Here a prime indicates the exclusion of the term $i=k$ from the product. So,  by applying Eqs. (\ref{eq:ADSTRANSF}) and (\ref{eq:ODDyalpha})  we write 
\begin{equation}
	\begin{aligned}
		&ds^2=-\frac{X}{U}\left[W dt-\sum_{i=1}^{n} a_i^2 \gamma_i d\varphi_i \right]^2+\frac{U}{X} dr^2+\sum_{\alpha=1}^{n-1} \frac{U_\alpha}{X_\alpha} dy_{\alpha}^2\\
		&+\sum_{\alpha=1}^{n-1}\frac{X_\alpha}{U_\alpha}\left[\frac{1+\frac{r^2}{l^2}}{1-\frac{y_{\alpha}^2}{l^2}} W dt - \sum_{i=1}^{n}\frac{a_i^2 (r^2+a_i^2)\gamma_i}{a_i^2-y_{\alpha}^2}d\varphi_i\right]^2\\
		&+\frac{\prod_{k=1}^{n} a_k^2}{r^2\prod_{\alpha=1}^{n-1}y_{\alpha}^2}\left[ (1+\frac{r^2}{l^2})Wdt-\sum_{i=1}^{n}(r^2+a_i^2)\gamma_i d\varphi_i\right] ^2,
	\end{aligned}
\end{equation}
where
\begin{equation}
	\begin{aligned}
		&U=\prod_{\alpha=1}^{n-1} (r^2+y_{\alpha}^2), \ \ \ U_{\alpha}=-(r^2+y_{\alpha}^2)\prod_{\beta=1}^{ n-1 \prime}  (y_{\beta}^2-y_{\alpha}^2), \ \ \ 1\leq \alpha \leq n-1, \\
		&W=\prod_{\alpha=1}^{n-1}(1-\frac{y_{\alpha}^2}{l^2}), \ \ \ \gamma_i=\frac{\prod_{\alpha=1}^{n-1} (a_i^2-y_{\alpha}^2)}{a_i \prod_{K=1}^{'n}(a_i^2-a_k^2) }, \ \ \ 1\leq i \leq n, \\
		&X= \frac{1+\frac{r^2}{l^2}}{r^2} \prod_{K=1}^{n}(r^2+a_k^2),\\
		&X_\alpha=\frac{1-\frac{y_\alpha^2}{l^2}}{y_\alpha^2}\prod_{K=1}^{n}(a_k^2-y_\alpha^2), \ \ \ 1\leq \alpha \leq n-1\, .
			\end{aligned}
	\end{equation}
Now by applying the same procedure as in earlier sections, we introduce the functions $f(r)$ and $g(r)$,
\begin{equation}
	\begin{aligned}
		&ds^2=\frac{f(r)}{U}\left[W dt-\sum_{i=1}^{n} a_i^2 \gamma_i d\varphi_i \right]^2-\frac{U}{g(r)} dr^2-\sum_{\alpha=1}^{n-1} \frac{U_\alpha}{X_\alpha} dy_{\alpha}^2\\
		&-\sum_{\alpha=1}^{n-1}\frac{X_\alpha}{U_\alpha}\left[\frac{1+\frac{r^2}{l^2}}{1-\frac{y_{\alpha}^2}{l^2}} W dt - \sum_{i=1}^{n}\frac{a_i^2 (r^2+a_i^2)\gamma_i}{a_i^2-y_{\alpha}^2}d\varphi_i\right]^2\\
		&-\frac{\prod_{k=1}^{n} a_k^2}{r^2\prod_{\alpha=1}^{n-1}y_{\alpha}^2}\left[ (1+\frac{r^2}{l^2})Wdt-\sum_{i=1}^{n}(r^2+a_i^2)\gamma_i d\varphi_i\right] ^2\, .
	\end{aligned} \label{eq:ADSMPFG}
\end{equation}
As before, by setting
\begin{equation}
		\begin{aligned}
	&y_\alpha^2=a_i^2, \ \ \  1\leq i \leq n-1, \ \ \  1\leq \alpha \leq n-1, \\
	&a_n=0,
		\end{aligned}
\end{equation}
we have
\begin{equation}
	ds^2=\frac{f(r)}{U}(W dt)^2-\frac{U}{g(r)} dr^2\, .
\end{equation}
Then, the Lagrangian,
\begin{equation}
	\mathcal{L}=\frac{f(r)}{U}W^2 \dot{t}^2-\frac{U}{g(r)} \dot{r}^2=\mu ^2,
\end{equation}
gives the conserved energy as
\begin{equation}
	p_t=\frac{f(r)}{U}W^2 \dot{t}=E,
\end{equation}
and leads  to 
\begin{equation}
	\dot{r}^2=\left(\frac{U}{f} W^2 E^2- \mu^2 \right) \frac{g}{U}\, .
\end{equation}
For a photon $(\mu=0)$, we write 
\begin{equation}
		\begin{aligned}
	&\ddot{r}=(\frac{g}{f})^{\prime} \ W^2 E^2=0 \Rightarrow \\
	& \frac{g}{f}=const=1\ .
\end{aligned}
\end{equation}
On the other hand, for $\mu=1$, we have
\begin{equation}
	\begin{aligned}
		&\dot{r}^2= W^2 E^2- \frac{f}{U} \Rightarrow \\
		&\ddot{r}=-\frac{1}{2}(\frac{f}{U})^{\prime}=-\frac{1}{2}(\frac{f^{\prime}}{U}-\frac{f U^{\prime}}{U^2}).
		\end{aligned}
\end{equation}
Writing
\begin{equation}
	f(r)=g(r)=X+\Psi(r),
\end{equation}
we have
\begin{equation}
	\ddot{r}=-\frac{1}{2}(\frac{X^{\prime}}{U}+\frac{\Psi^{\prime}}{U}-\frac{XU^{\prime}}{U^2}-\frac{\Psi
		U^{\prime}}{U^2}),
\end{equation}
which should be equal to the Newtonian acceleration far from a black hole, and for very large values of $r$, $U\approx r^{2n-2}, X\approx r^{2n-2}+\frac{r^{2n}}{l^2}$,  we have
\begin{equation}
\ddot{r}=-\frac{1}{2}\left( \frac{\Psi^\prime}{r^{2n-2}}+\frac{2r}{l^2}-\frac{(2n-2)\Psi}{r^{2n-1}}\right)\, = -\frac{(3-d)M}{r^{d-2}}-\frac{r}{l^2}\, . 
\end{equation}
This is then integrated to give
\begin{equation}
	\Psi(r)=-2M
\end{equation}
and 
\begin{equation}
	f(r)=g(r)=X-2M\, .
\end{equation} 
Putting this into the metric (\ref{eq:ADSMPFG})  would describe a rotating black hole in AdS-dS ($AdS \rightarrow dS$ by $1/l^2 \rightarrow - 1/l^2$) spacetime in arbitrary odd dimensions$(d=2n+1)$.

For even $(d=2n+2)$ dimensions, we can follow the same procedure, beginning with the metric (where there is an unpaired coordinate $z$),  
\begin{equation}
	ds^2=-\left(1+\frac{\sum_{i=1}^{n}({x}_i^2+{y}_i^2)+z^2}{l^2}\right) dt^{\prime2}+dz^2+\sum_{i=1}^{n} ({dx}_i^2+{dy}_i^2)-\frac{\left( z^2 dz^2+\sum_{i=1}^{n} ({x}_i dx_i+{y}_i dy_i)\right) ^2}{l^2+z^2+\sum_{i=1}^{n} ({x}_{i}^{2}+{y}_{i}^{2})}
\end{equation}
and the transformations
\begin{equation}
	\begin{aligned}
		&x_i=\sqrt{\frac{(r^2+a_i^2)}{1-\frac{a_i^2}{l^2}}} \ \mu_i \  cos[(1-\frac{a_i^2}{l^2})\varphi_i]\\
		&y_i=\sqrt{\frac{(r^2+a_i^2)}{1-\frac{a_i^2}{l^2}}} \ \mu_i \  sin[(1-\frac{a_i^2}{l^2})\varphi_i]\\
		&z=r \alpha,\\
		&t^{\prime}=\prod_{i=1}^{n}(1-\frac{a_i^2}{l^2})t,
	\end{aligned}
\end{equation}
where
\begin{equation}
	\sum_{i=1}^{n}{\mu_i}^2+\alpha^2=1\, .
\end{equation} \\

By retracing the same steps, one would easily obtain the corresponding metric for even $(d=2n+2)$ dimensions. By setting $1/l^2=0$, we recover the Myers-Perry solution for arbitrary dimension $d$.\\

\section{\bfseries Discussion}

The main purpose of both the NJ algorithm \cite{kerrnewman1} and Dadhich's method \cite{Dadhich} is to obtain the rotating black hole metric without having to solve the formidable Einstein vacuum equations. Both techniques work wonderfully well. The working of the former is however not so clear and transparent. In contrast, the latter is driven  entirely by the two physically motivated  guiding conditions: (a) a
photon experiences no acceleration, so as to keep the
velocity of light constant (this condition is equivalent to the
null energy condition), and (b) a timelike particle experiences Newtonian acceleration in the first approximation so as to include Newtonian gravity in general relativity \cite{DadhichSchw}. These are the two natural features that should be incorporated as one goes from Newtonian theory to general relativity. This is therefore the most attractive and physically satisfying aspect of Dadhich's prescription.\\

The NJ algorithm has also been applied in modified gravity \cite{NJ2}; however, it does not always work \cite{DadhichGosh}, and it has been strongly argued that it should not be applied outside general relativity \cite{NJ2}. Its application in generating higher-dimensional rotating black holes also does not go beyond $d=5$ and $d=7$ (equal angular momenta) \cite{Mirza5D,MirzaOdd}. \\

On the other hand, Dadhich's method \cite{Dadhich} which we have employed in this paper to obtain a Myers-Perry rotating black hole in arbitrary dimensions in AdS-dS spacetime works wonderfully well in arbitrary dimensions, and $\Lambda$ could also be easily included. In a similar manner, it was also applied to write the pure Gauss-Bonnet rotating black hole metric in six dimensions. It does however lead to a perfectly acceptable black hole metric with all of the expected features of a six-dimensional pure Gauss-Bonnet black hole \cite{DadhichGosh}. Unfortunately, this is not an exact solution of the pure Gauss-Bonnet vacuum equation, which is satisfied only at leading orders. This  may be because of the pure Gauss-Bonnet equation is quadratic in Riemann curvature, and hence what worked for general relativity for which the equation was linear in Riemann may not work when it is nonlinear in Riemann. However, it works for arbitrary dimension with the inclusion of $\Lambda$. As
we have noted that NJ algorithm has not been applied for
$d > 7$, and in contrast Dadhich’s method \cite{Dadhich} works
beautifully in any arbitrary dimension as we have shown
in obtaining the higher-dimensional Myers-Perry rotating
black hole metric. \\ 
\section{\bfseries Acknowledgment}
N. D. wishes to warmly acknowledge the support of
CAS President’s International Fellowship Initiative Grant
No. 2020VMA0014.\\


\end{document}